\documentclass{ws-mpla}

\usepackage{xcolor}
\usepackage{graphicx}
\usepackage[verbose,hypertexnames=false]{hyperref}
\hypersetup{colorlinks=false,allbordercolors=blue,pdfborderstyle={/S/U/W 1}}

\begin{document}

\markboth{Ruo-Yu Liu}{PeV Gamma-Ray Emitters: Observations and Implications}

\catchline{}{}{}{}{}

\title{PeV Gamma-Ray Emitters: Observations and Implications}

\author{Ruo-Yu Liu}

\address{School of Astronomy and Space Science, Nanjing University, 163 Xianlin Avenue, \\
Nanjing 210023, China\\
ryliu@nju.edu.cn}

\maketitle

\begin{history}
\received{(Day Month Year)}
\revised{(Day Month Year)}
\accepted{(Day Month Year)}
\published{(Day Month Year)}
\end{history}

\begin{abstract}
The origin of Galactic cosmic rays (CRs) around and beyond the knee remains an open problem. Gamma rays extending to the PeV range provide direct evidence for extreme particle acceleration. 
LHAASO has greatly expanded the observational sample of such sources. In this review, we use the term \emph{PeV gamma-ray emitter} for a source whose measured spectrum extends to, or approaches, 1~PeV, rather than restricting it to sources detected above 1~PeV at more than $5\sigma$. We review LHAASO observations of six PeV gamma-ray emitters (i.e., the Crab Nebula, the Aquila Booster, the Cygnus Bubble, Cygnus~X-3, V4641~Sgr, and GRS~1915+105) together with relevant multiwavelength observations and on-going theoretical studies. Together, these sources show that PeV particle acceleration can occur in different environments and on spatial scales ranging from compact pulsar-wind nebulae and binary-scale jets to large-scale jet--environment interaction regions and extended star-forming complexes. Their spectra, morphologies, variability, magnetic fields, and surrounding gas provide important constraints on particle acceleration and transport mechanisms. We also discuss the implications of these observations for the origin of Galactic PeV CRs and future observational prospects for clarifying this problem.
\end{abstract}


\ccode{PACS Nos.: 03.65.$-$w, 04.62.+v}

\section{Introduction}\label{sec:introduction}

The all-particle cosmic-ray (CR) spectrum extends over many decades in energy and contains several structures that may reflect changes in acceleration, transport, or transitions of the dominant source population. The most famous feature caused by Galactic CR sources is the so-called ``knee'' beyond which the spectrum becomes noticeably steeper. Early air-shower experiments, including KASCADE, AS$\gamma$, and IceTop, found the knee at an energy around a few PeV, although the reported knee position and absolute flux differed because of uncertainties in the energy scale, mass composition, and hadronic interaction models \cite{antoni2005kascade,amenomori2008tibetknee,aartsen2019icetop}. More recently, LHAASO measured the all-particle spectrum and the mean logarithmic mass from 0.3 to 30~PeV with much improved precision. It placed the knee at \(3.67\pm0.05_{\rm stat}\pm0.15_{\rm sys}\)~PeV and found that the composition becomes lighter below the knee and heavier above it, showing that light nuclei make an important contribution to the knee \cite{lhaaso2024allparticle}. The separate proton and helium measurements provide a clearer picture. The proton spectrum from 0.15 to 12~PeV shows a strong hardening below the knee. This hardening reaches its maximum at about 3~PeV and is followed by a sharp softening near 3.3~PeV \cite{lhaaso2025proton}. On the other hand, the helium spectrum from 0.16 to 13~PeV hardens around 1.1~PeV and softens near 7~PeV, forming a helium knee \cite{lhaaso2026helium}. Interestingly, helium is softer than protons over much of the PeV regime, with protons overtaking helium at about 0.7~PeV while helium becomes dominant again at about 5~PeV. The different spectral shapes and the strong energy dependence of the proton-to-helium ratio are difficult to explain with a single source population. They suggest that several Galactic source populations, with different elemental abundances and maximum rigidities, contribute to the CR spectrum around and beyond the knee.

As charged particles, CRs cannot be traced directly back to their sources because their directions are strongly deflected by intervening magnetic fields between their sources and the Earth. We have to locate their accelerators through neutral messengers, mainly gamma rays and neutrinos produced by the accelerated particles. Therefore, ultrahigh-energy (UHE, photons of energy $E>100\,$TeV) gamma-ray sources are candidate sources of PeV particle accelerators. LHAASO opened this observational window with more than 530 photons above 100\,TeV from twelve Galactic sources, collected by its half array in the first 11 months \cite{lhaaso2021_12sources}, including events up to 1.4\,PeV. Its first catalog reported 43 UHE sources at more than $4\sigma$ in our Galaxy. These UHE sources contain several types of Galactic objects, including pulsar wind nebulae (PWNe), young massive star clusters, microquasars, and supernova remnants (SNRs) \cite{cao2021uhe,cao2024catalog,deona2024}. SNRs have long been regarded as the standard sources of Galactic cosmic rays, but they are not the dominant population in the UHE band. Several well-studied SNRs are either not detected in the UHE band or show gamma-ray spectra that have already softened or entered the cutoff regime before the UHE band \cite{magic2017casa,lhaaso2024casa,lhaaso2026ic443,veritas2017tycho,hess2007rxj1713,hess2018velajr}. In contrast, several PWNe and microquasars show relatively hard spectra extending to several hundred TeV or even into the PeV regime. These observations show that PWNe and microquasars constitute important classes of extreme particle accelerators, although their contribution to the Galactic CR population around and beyond the knee remains uncertain.

The origin of the UHE gamma rays can be either hadronic or leptonic. Hadronic interactions of protons or nuclei produce gamma rays through neutral-pion decay, while electrons can produce gamma rays through inverse-Compton (IC) scattering. 
However, a leptonic-dominated gamma-ray spectrum does not imply that the source is unable to accelerate protons. Hadronic gamma-ray production requires not only energetic protons, but also sufficient target material for efficient pion production. A source may contain PeV protons and still be faint in hadronic gamma rays if the gas and radiation density is low or if the particles escape rapidly. Moreover, accelerating electrons to PeV energies is a strong indication of extreme acceleration, because electrons suffer severe radiative losses, particularly through synchrotron radiation. If protons or nuclei are injected into the same acceleration zone, their much weaker radiative losses should allow them to reach at least comparable or often higher energies. 

So far, we have not yet firmly identified sufficient PeVatrons to account for the measured proton flux around and beyond the knee. Therefore, at the current stage, it is important to ask what kinds of Galactic objects can sustain particle acceleration up to the energies relevant to the CR knee. Gamma rays approaching 1~PeV provide a straightforward way to address this question. In a leptonic scenario, their production requires electrons with energies of a few PeV despite their rapid radiative cooling. In a hadronic scenario, they point to protons or nuclei reaching several PeV or even higher. Although the radiation mechanism must be determined source by source, a gamma-ray spectrum extending close to the PeV range already provides strong evidence for an extreme accelerator.
Therefore, this review will mainly focus on PeV gamma-ray emitters detected by LHAASO so far. The term ``PeV gamma-ray emitter'' here describes sources whose spectra extend to, or approach, PeV energies, rather than restricting it to sources detected above 1~PeV at more than $5\sigma$\footnote{As of now, only Cygnus~X-3 meets this strict definition.}. It then includes two PWNe (the Crab Nebula and the Aquila Booster), one extended intense star-forming environment (the Cygnus Bubble), and three microquasar systems (V4641~Sgr, GRS~1915+105 and Cygnus~X-3), where the relevant acceleration sites range from compact nebulae and binary-scale jets to extended regions many parsecs away from the central engine. These sources may not all contribute equally to the CR flux measured at Earth, because their particle content, energy budget, and escape histories can be very different. Nevertheless, taken together, they provide a new observational sample with which we can study how, rather than whether, particles can be accelerated to the PeV regime. We will review the LHAASO observations of these sources together with relevant measurements at other wavelengths, as well as our current understanding of these objects. By comparing these physically different systems, we will try to clarify where and how the highest-energy particles are accelerated, how they escape and produce observable radiation, and what these results imply for the origin of Galactic CRs around and beyond the knee.

\section{Instruments and Observations}
\label{sec:instrument}

Located at an altitude of 4410~m, LHAASO contains two main arrays for gamma-ray observations \cite{lhaaso_wb}. The 78,000-m\(^2\) Water Cherenkov Detector Array (WCDA) covers the sub-TeV to multi-TeV range and provides an important connection to measurements by conventional TeV instruments, while the 1.3-km\(^2\) Kilometer Square Array (KM2A) is optimized for gamma rays above $\sim10$~TeV and extends the measurement into the PeV regime. KM2A consists of a large array of 5216 electromagnetic particle detectors together with 1188 underground muon detectors, which provide efficient rejection of the much more abundant hadronic CR showers. Its effective area approaches a square kilometer at high energies, and the angular resolution improves from about $0.4^\circ$ at 30~TeV to about $0.2^\circ$ at PeV energies \cite{lhaaso_wb}. Together with its large field of view and high duty cycle, these properties make LHAASO highly sensitive to UHE gamma rays and enable the detection of PeV gamma-ray photons.

The interpretation of measurements near the PeV range nevertheless requires some care. Generally, the gamma-ray flux of a source decreases rapidly with energy, and even for the brightest sources only a small number of photons are detected in the highest-energy part of the spectrum. Consequently, the high-energy spectrum tail is generally determined through a likelihood analysis in which the detector response is folded with an assumed spectral form and spatial model, rather than through a completely model-independent measurement of the flux in each narrow energy bin. When the number of events becomes small, the inferred flux and spectral cutoff can therefore depend more strongly on the adopted spectral and morphological templates. This is particularly important in crowded regions of the Galactic plane. Emission from neighboring sources, unresolved sources, and Galactic diffuse emission can contribute photons within the same region of interest, and their separation depends on the assumed spatial templates and on the energy-dependent point-spread function. The problem can become more complicated if the morphology of an extended source changes with energy. The diffuse Galactic gamma-ray emission (DGE) has now been measured by LHAASO up to the PeV range and is therefore not negligible when extended UHE sources are studied \cite{lhaaso2023diffuse,lhaaso2025diffuse}. On the other hand, photons around PeV energies suffer absorption by the interstellar radiation field (ISRF) and the cosmic microwave background (CMB) through $\gamma\gamma\rightarrow e^{+}e^{-}$ during propagation. Absorption by infrared photons in the ISRF can already become important around $100$--$200$~TeV for distant sources, while the CMB increasingly dominates above several hundred TeV \cite{vernetto2016absorption}. The observed spectrum can therefore be substantially softer than the intrinsic one for sources located several kiloparsecs away. Recovering the intrinsic spectrum requires a correction that depends on both the source distance and its position in the Galaxy. Uncertainties in the distance and in the Galactic radiation field would propagate directly into the inferred intrinsic spectrum, cutoff energy, and luminosity of a PeV gamma-ray source.

For these reasons, LHAASO measurements become much more informative when combined with observations at lower energies and at other wavelengths. GeV observations with \textit{Fermi}-LAT constrain the low-energy gamma-ray spectrum and can reveal whether the UHE emission connects smoothly to the GeV component or requires an additional population of particles. Imaging atmospheric Cherenkov telescopes such as H.E.S.S., MAGIC, and VERITAS have much better angular resolution in the TeV range and can resolve structures that are difficult to separate with an air-shower array. Their measurements are useful for identifying the morphology and possible counterparts of LHAASO sources, while measurements by HAWC and AS$\gamma$ provide independent coverage in the multi-TeV range. X-ray and radio observations provide another important constraint as synchrotron radiation probes the relativistic electron population and magnetic field. X-ray morphology can in particular trace the highest-energy electrons and their cooling close to the acceleration region, whereas radio observations probe lower-energy electrons over larger spatial scales. Measurements of atomic and molecular gas through H~I and CO surveys are essential for testing hadronic interpretations, since the expected gamma-ray luminosity depends directly on the amount and distribution of target gas. Finally, a coincident neutrino signal would provide the smoking gun for hadronic interactions. The nature of a PeV gamma-ray emitter therefore cannot usually be established from its highest-energy photons alone. Its spectrum, morphology, surrounding environment, and multiwavelength and multimessenger counterparts must be considered together.

\section{PeV gamma-ray emitters}
\label{sec:pev-emitters}

The PeV gamma-ray emitters discussed here are powered by different central engines and produce radiation on different spatial scales. In general, they can be categorized into three source classes: (1) pulsar wind nebulae (PWNe), which test how efficiently the rotational energy of a neutron star can be converted into PeV particles; (2) microquasars, which probe particle acceleration in black-hole jets and their interactions with surrounding medium; (3) the young massive star cluster in which massive stars and supernovae continuously inject mechanical energy into surrounding medium.

\subsection{Pulsar wind nebulae}
\label{sec:pwn}

PWNe are among the most prominent source classes in the UHE gamma-ray sky. A rapidly rotating pulsar launches a relativistic wind dominated by electrons and positrons. The wind is decelerated at a termination shock and transfers its energy to the surrounding nebula. The maximum particle energy achievable in a PWN is related to the pulsar's spindown power $L_{\rm sd}$ \cite{deJager1992,amato2021,emma2022pwn}. If a fraction $\epsilon_B$ of the total power (i.e., the spindown power in the case of pulsar) is carried by the magnetic field, the magnetic field at the acceleration zone satisfies
\begin{equation}\label{eq:B_pwn}
    \frac{B_{\rm acc}^2}{8\pi}
    =\frac{\epsilon_BL_{\rm sd}}{4\pi  R_{\rm acc}^2\beta c},
\end{equation}
where $R_{\rm acc}$ is the radius of the acceleration zone {  and $\beta$ is the speed of the pulsar wind in unit of the speed of light.  Since pulsar wind is generally ultra-relativistic, we may take the approximation $\beta\simeq 1$.} Writing the effective accelerating electric field as $\mathcal{E}=\eta B_{\rm acc}$ \cite{aharonian2002}, the maximum energy of a particle with charge $Z$ allowed by the available potential is
\begin{equation}
 E_{\max}
 \simeq \eta Ze B_{\rm acc}R_{\rm acc}
 =\eta Ze\epsilon_B^{1/2}
 \left(\frac{2L_{\rm sd}}{c}\right)^{1/2}
 \simeq
 7.8\,\eta Z\epsilon_B^{1/2}
 \left(\frac{L_{\rm sd}}{10^{37}\,\rm erg/s}\right)^{1/2}\,{\rm PeV},
 \label{eq:pwn_emax}
\end{equation}
We see that $E_{\max}\propto L_{\rm sd}^{1/2}$ and the characteristic size of the acceleration zone cancels out. In the most optimistic limit (i.e., $\eta=1$), this energy is comparable to the full potential drop available in the pulsar magnetosphere, which also corresponds to the maximal limit under the classical electrodynamics and ideal magnetohydrodynamic (MHD) condition. For electrons, the actual maximum energy can be lower because synchrotron and IC losses compete with acceleration. If synchrotron losses dominate, the maximum energy can be given by 
\begin{equation}
    E_{\rm max}=6.3\,\eta^{1/2}\epsilon_B^{-1/2}\left(\frac{R_{\rm acc}}{\rm 0.1\,pc}\right)^{1/2}\left(\frac{L_{\rm sd}}{10^{37}\,\rm erg/s}\right)^{-1/4}\,\rm PeV.
    \label{eq:pwn_max_c}
\end{equation}
Depending on the magnetization and size of a pulsar, the maximum electron energy is the smaller of Eq.~(\ref{eq:pwn_emax}) and Eq.~(\ref{eq:pwn_max_c}).

\begin{figure}
    \centering
    \includegraphics[width=0.7\linewidth]{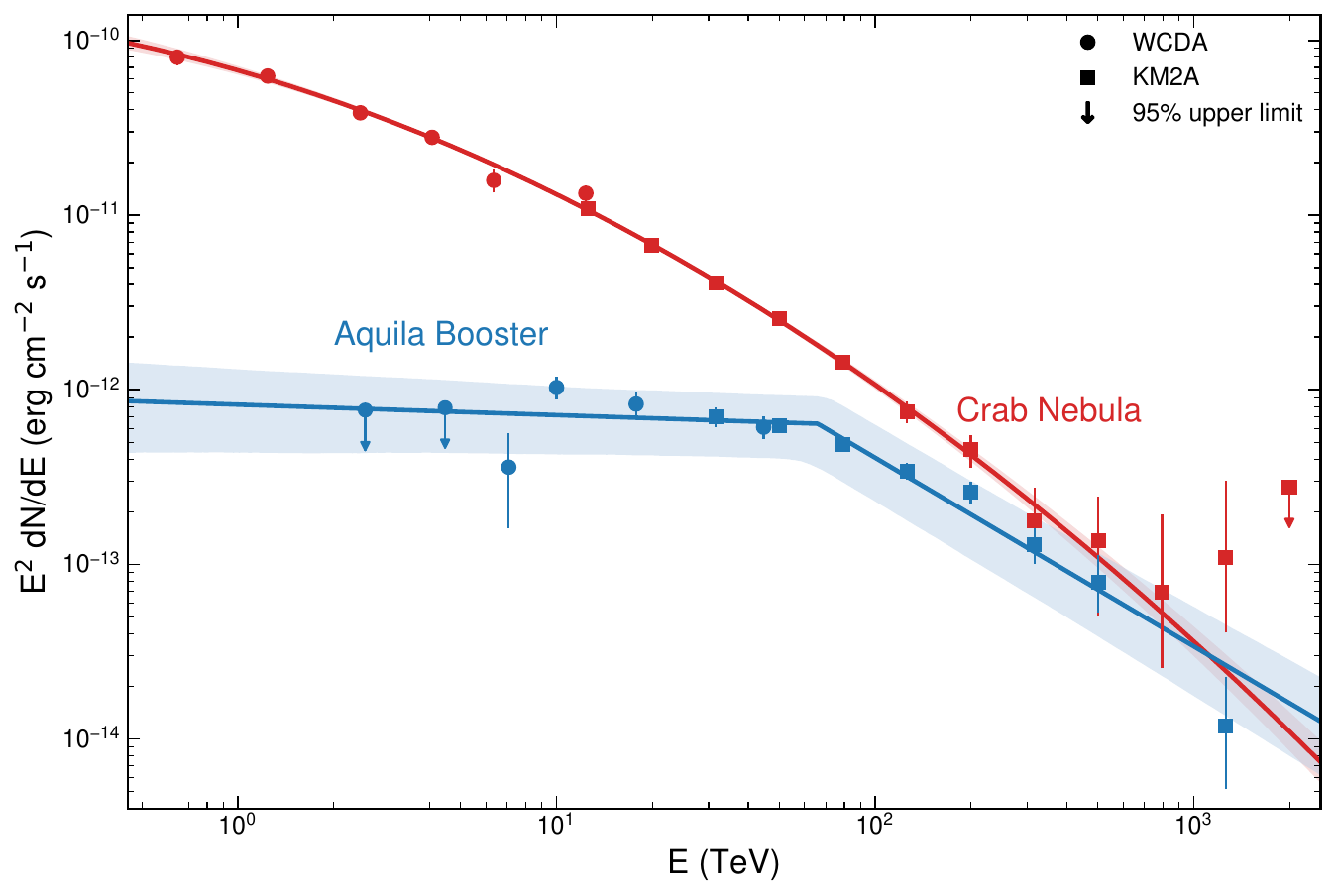}
    \caption{Spectral energy distributions of the Crab Nebula (red) and the Aquila Booster associated with PSR J1849$-$0001 (blue), as measured by LHAASO. Circles and squares denote measurements with WCDA and KM2A, respectively, and downward arrows indicate 95\% confidence-level upper limits. The solid curves show the best-fit spectral models, with the shaded regions indicating the corresponding uncertainties. The Crab Nebula data are taken from the LHAASO measurement reported in Ref.\protect\cite{cao2021crab}, while the Aquila Booster data are taken from Ref.\protect\cite{lhaaso2026aquila}.}
    \label{fig:pwn}
\end{figure}

\subsubsection{The Crab Nebula}
The Crab Nebula was the first astrophysical source discovered at TeV band \cite{Weekes1989}. It was also the first source firmly detected above 100~TeV. AS$\gamma$ reported a significant signal beyond 100~TeV, and HAWC measured its spectrum in the same energy range \cite{amenomori2019crab,abeysekara2019crab}. These observations had already implied electrons approaching the PeV range and placed strong constraints on the magnetic field and acceleration conditions in the nebula \cite{khangulyan2020crab,zhang2020crab}. LHAASO subsequently extended the gamma-ray spectrum to about 1.1~PeV (Fig.~\ref{fig:pwn}), which requires electrons to be accelerated at least to 2\,PeV under the IC interpretation \cite{cao2021crab}.

The spindown power of the Crab pulsar is $L_{\rm sd}\simeq4.5\times10^{38}\,{\rm erg\,s^{-1}}$. Eq.~(\ref{eq:pwn_emax}) gives an ideal potential-limited energy of order 50~PeV for a singly charged particle if $\eta$ and $\epsilon_B$ are close to unity. The 2\,PeV electrons required by the LHAASO observation are still below the absolute electrodynamic and ideal-MHD limit, but producing them is still difficult because of their rapid synchrotron cooling. From the multiwavelength SED modeling, the LHAASO analysis inferred an average magnetic field of order $100\,\mu{\rm G}$, implying an acceleration efficiency of about $\eta>0.16$\cite{cao2021crab}.

The physical mechanism that can accelerate electrons to such a high energy remains uncertain. The pulsar-wind termination shock is the most commonly considered site, but standard diffusive shock acceleration is inefficient at a highly relativistic, quasi-perpendicular shock, especially when the upstream wind remains significantly magnetized \cite{sironi2009shock}. Several mechanisms have been discussed. In the striped equatorial wind, compression of the alternating magnetic field at the termination shock can trigger magnetic reconnection and efficiently transfer Poynting flux to nonthermal particles \cite{lyubarsky2003striped,sironi2011stripedwind}. The global geometry of the termination shock may also be important. Close to the equatorial current sheet, field reversals and particle drifts can allow repeated shock crossings and efficient Fermi-like acceleration even when local plane-shock calculations predict inefficient acceleration \cite{giacinti2018crab,cerutti2020global}. Magnetic reconnection and turbulence inside the nebula provide additional channels for rapid particle energization and are independently motivated by the Crab gamma-ray flares, which demonstrate that non-ideal electric fields can accelerate electrons beyond the classical synchrotron burn-off limit \cite{uzdensky2011crabflare,cerutti2013crabflare,sironi2014reconnection}. However, the compact flaring regions are not necessarily the same sites that produce the steady PeV IC emission. Stochastic acceleration by nebular turbulence has also been studied and may help shape the broad particle spectrum, although it is less clear whether it can reach the highest electron energies before cooling becomes dominant \cite{tanaka2017crab,tanaka2024stochastic}.

On the other hand, although the overall broadband emission is well described by leptonic models, the LHAASO flux around 1~PeV shows a possible excess relative to the smoothly steepening one-zone inverse-Compton spectrum \cite{cao2021crab}. The significance is not sufficient to claim an additional component, but LHAASO showed that the PeV flux may also be reproduced by $\pi^0$-decay emission from protons. In principle, the pulsar wind may contain an ion component, and these ions can be accelerated together with electrons/positrons \cite{hoshino1992ion,atoyan1996crab,amato2006ion}. For an average target gas density of $n\simeq10\,{\rm cm^{-3}}$, the required energy in confined protons is about $3\times10^{47}$~erg for an $E_p^{-2}$ spectrum with a cutoff at 30~PeV, or about $5\times10^{46}$~erg for a narrow proton component around 10~PeV \cite{cao2021crab}. Because the $pp$ cooling time is very long in the Crab Nebula, the required proton injection power depends strongly on how efficiently these particles are confined. Once the escape of relativistic protons is included, the total energy injected into ions over the lifetime of the nebula may be much larger than the amount remaining there today, which would imply the Crab Nebula to be an efficient PeV proton accelerator \cite{liu2021crab}. Spatially dependent lepto-hadronic models have shown that a proton component could dominate the emission above a few hundred TeV under suitable conditions \cite{peng2022crab,liang2025pwn}, and the proton component can carry $\lesssim 5\%$ of the spindown power of the Crab pulsar\cite{liang2025pwn}. Re-acceleration of ions from the supernova ejecta at the termination shock provides another possible origin for such hadrons \cite{spencer2025crab}. The present data do not require a hadronic component, but neither do they exclude the hadronic origin of the highest-energy photons which are relevant to the Galactic CR knee. Detection of photons well beyond a few times PeV, or an associated neutrino signal, would provide a much stronger test of this possibility.

\subsubsection{The Aquila Booster}
Aquila Booster is a PWN powered by PSR~J1849-0001 located at about 7\,kpc away from Earth \cite{Gotthelf2011}. H.E.S.S. and AS$\gamma$ detected gamma rays from this source up to about 320~TeV \cite{hess2018_survey,amenomori2023j1849}, while X-ray observations revealed a compact PWN and showed that its magnetic field is unusually weak \cite{Gagnon2024,kim2024j1849}. LHAASO detected a point-like UHE source with a hard spectrum extending to 2\,PeV \cite{lhaaso2026aquila} (Fig.~\ref{fig:pwn}). Interestingly, the spindown luminosity of PSR~J1849-0001 is only $L_{\rm sd}\simeq9.8\times10^{36}\,{\rm erg\,s^{-1}}$, about 50 times lower than that of the Crab pulsar, whereas its PeV gamma-ray luminosity is several times higher. A joint fit to the X-ray synchrotron and gamma-ray IC emission gives an average magnetic field of only $B=2.8\pm0.2\,\mu{\rm G}$, almost two orders of magnitude below that in the Crab \cite{lhaaso2026aquila}. Since the relative synchrotron and IC losses scale approximately as $U_B/U_{\rm ph}$, the weak magnetic field allows a much larger fraction of the electron energy to go into gamma rays via the IC radiation. 

However, this is not a favorable condition for accelerating electrons to 2\,PeV, which is needed to account for the 2\,PeV photon. Because the spindown power of J1849 is about 50 times lower than that of the Crab, its available potential is smaller by a factor of about seven. To accelerate electrons to 2~PeV, the potential-drop constraint alone requires $\eta \gtrsim 0.27\,\epsilon_B^{-1/2}$. Even in the extreme limit $\epsilon_B=1$, the required efficiency is already more demanding than that of the Crab Nebula. The situation becomes much more severe if acceleration occurs at the termination shock. Although the magnetic field inferred from the multiwavelength SED modeling is volume averaged and does not directly measure the field at the shock, both the energy budget and previous multi-zone modeling suggest a small magnetic-energy fraction, $\epsilon_B\sim 10^{-3}-10^{-2}$, near the termination shock \cite{kim2024j1849,lhaaso2026aquila}. With such a low magnetization, accelerating 2~PeV electrons at the termination shock would formally require $\eta> 1$. 

This result suggests that the PeV particles may not be accelerated by an ordinary ideal-MHD process at the termination shock. One possibility is that acceleration occurs farther upstream, before most of the wind Poynting flux has been converted or dissipated. The striped pulsar wind can contain a much larger magnetic energy fraction at smaller radii, and magnetic reconnection or other non-ideal electric fields may tap a larger fraction of the pulsar electromotive potential \cite{lyubarsky2001striped,aharonian2012coldwind,lhaaso2026aquila}. The precise location of the accelerator is still uncertain, but the Aquila Booster provides an indication that the highest-energy particles may be energized upstream of the conventional termination shock.

Note that changing the interpretation of the highest-energy spectrum from leptonic to hadronic does not solve the problem. Radiative cooling is much weaker for protons, but cooling is not the dominant constraint for the Aquila Booster. Producing gamma rays near 2~PeV through $pp$ interactions would require protons with energies of about 20~PeV. Since the available electric potential depends on particle charge rather than mass, a hadronic interpretation would therefore make the electrodynamic requirement even more demanding. LHAASO's observation of Aquila Booster does not simply reveal another PWN that contains PeV particles, but suggests that the extreme particle acceleration efficiency is not unique to the Crab Nebula and could be common among PWNe. If so, young PWNe of relatively high spindown power might serve as a non-negligible population of PeV CR contributors, provided that sufficient protons/ions are loaded into the acceleration zones.

\subsection{Microquasars}
\label{sec:microquasars}

Microquasars have only recently been discovered in the UHE band \cite{lhaaso2025microquasars,hawc2024v4641}. Accretion onto a stellar-mass black hole powers a relativistic jet, and particles may be accelerated close to the compact object, within the accretion disk or propagating jet, or where the jet interacts with the surrounding medium. An important result from LHAASO is that these different spatial scales are now directly visible in the gamma-ray data. LHAASO detected gamma-ray sources associated with five black hole--jet systems: SS~433, V4641~Sgr, GRS~1915+105, MAXI~J1820+070, and Cygnus~X-1 \cite{lhaaso2025microquasars,lhaaso2026grs1915}. In particular, the spectrum of V4641~Sgr and GRS~1915+105 extends to approximately 1\,PeV. More importantly, the morphology generally does not point to the immediate vicinity of the black hole. SS~433, V4641~Sgr, and GRS~1915+105 show extended UHE emission. MAXI~J1820+070 is consistent with a point source, but its centroid is displaced by about $0.27^\circ$ from the black hole along the direction of the receding radio ejecta. A smaller offset is also seen in Cygnus~X-1, approximately along the radio bow structure inflated by its jet \cite{lhaaso2025microquasars}. These observations suggest that UHE emissions from these sources are mainly produced after energetic particles have propagated away from the binary and interacted with the large-scale jet or the surrounding interstellar medium.

\begin{figure}
    \centering
    \includegraphics[width=0.7\linewidth]{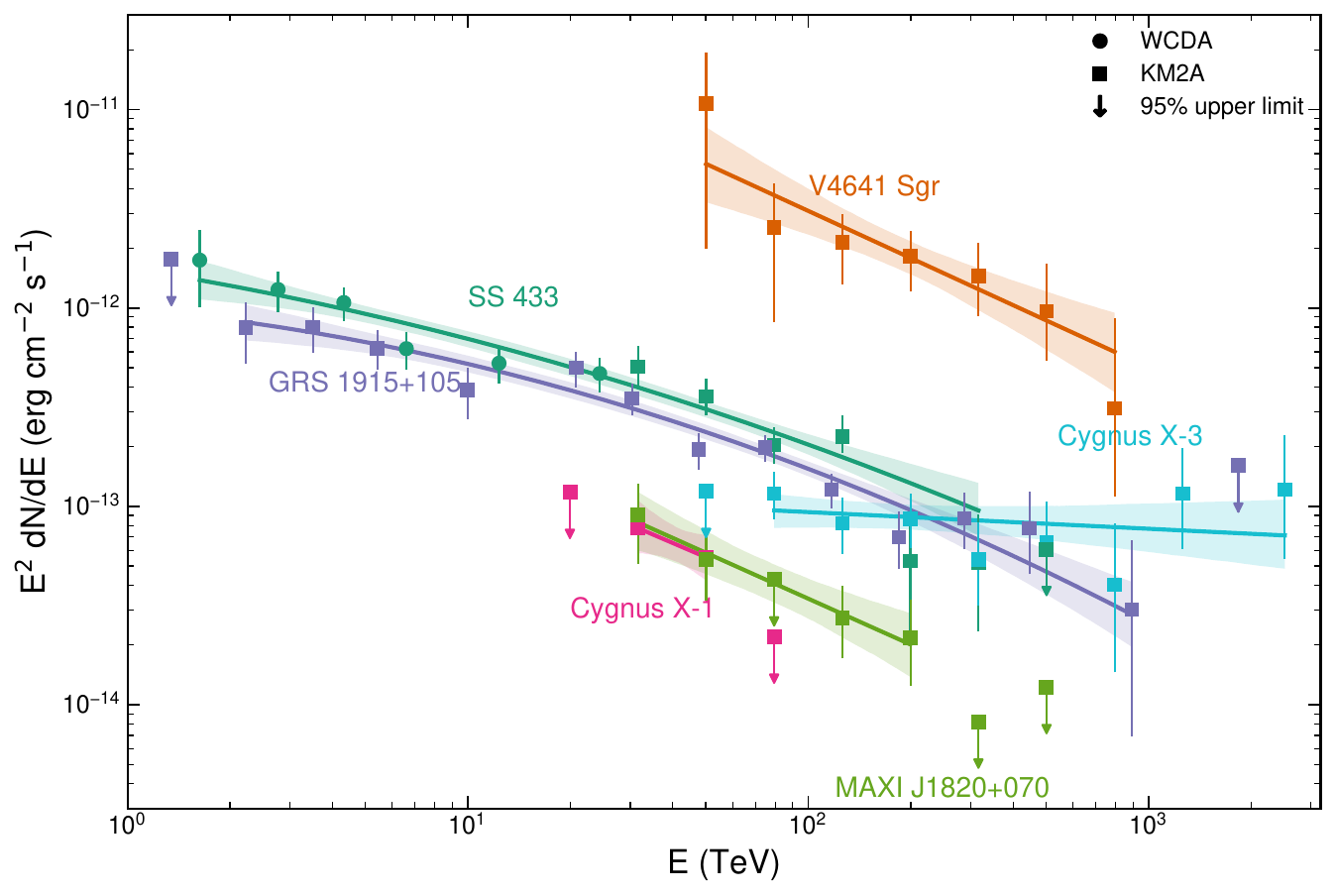}
    \caption{ Spectral energy distributions of six Galactic microquasars measured by LHAASO: SS~433, V4641~Sgr, GRS~1915+105, MAXI~J1820+070, Cygnus~X-1, and Cygnus~X-3. Different colors represent different systems. Circles and squares denote WCDA and KM2A measurements, respectively, and downward arrows indicate 95\% confidence-level upper limits. For SS~433, the fluxes associated with the eastern and western lobes are summed below 100~TeV, whereas the emission above 100~TeV includes the central UHE component. The solid curves and shaded regions show the best-fit function for spectra and their corresponding uncertainties. The measurements of SS~433, V4641~Sgr, MAXI~J1820+070, and Cygnus~X-1 are taken from Ref.\protect\cite{lhaaso2025microquasars}. For GRS~1915+105, we use the updated LHAASO spectrum reported in Ref.\protect\cite{lhaaso2026grs1915}, while the Cygnus~X-3 spectrum is taken from Ref.\protect\cite{lhaaso2026cygx3}.}
    \label{fig:microquasar}
\end{figure}

On the other hand, LHAASO detection of Cygnus~X-3 reveals an opposite scenario. Unlike the five systems above, its PeV emission is variable on month timescales and shows indication of modulation at the 4.8-h orbital period \cite{lhaaso2026cygx3}. Therefore, the emitting region must lie within or close to the binary. Microquasars can accelerate PeV particles over a wide range of spatial scales, from the {  vicinity of black-hole environment/jet base} to jet--ISM interaction regions tens or even hundreds of parsecs away. TeV-PeV spectra of these microquasars are compared in Fig.~\ref{fig:microquasar}

{  Similar to the PWN case, the maximum achievable particle energy in microquasars is also limited by the available potential in the acceleration zone. The particle acceleration in microquasars is usually considered to be relevant with the accretion-powered outflow, either jet \cite{WangJS2025} or disk wind \cite{Peretti2025}. Denoting the kinetic power of the outflow by $L_{\rm k}$, we may estimate the magnetic field in the acceleration zone through
\begin{equation}
    \frac{B^2_{\rm acc}}{8\pi}=\frac{\sigma L_{\rm k}}{\Omega R^2_{\rm acc}\beta c},
\end{equation}
where $\Omega$ is related to the geometry of the acceleration zone. For an isotropic outflow, $\Omega=4\pi$ and $R_{\rm acc}$ is the distance of the acceleration zone from the compact object, while for a conical jet $\Omega=\pi$ and $R_{\rm acc}$ is the transverse radius of the acceleration zone. There are a few differences from the case of PWN (Eq.~\ref{eq:B_pwn}). First, the outflow in a microquasar is not ultra-relativistic, so we cannot simply employ $\beta\simeq 1$. The effective accelerating electric field correspondingly becomes $\mathcal{E}=\eta \beta B_{\rm acc}$. In addition, the kinetic energy of an outflow in a microquasar is not necessarily the total energy budget of the system, we here use the magnetization factor $\sigma$, which is defined as the ratio between the magnetic power and the kinetic power, to calculate the magnetic field. If the outflow is Poynting flux dominated, the magnetic power may be higher than the kinetic power (i.e., $\sigma>1$). Provided that the total energy budget of the outflow before particle acceleration is mainly composed of kinetic energy and magnetic energy, we have $\sigma=\epsilon_{B}/(1-\epsilon_B)$. , we may estimate the maximum energy to be 
\begin{equation}
 E_{\max}
 \simeq \eta Ze \beta B_{\rm acc}R_{\rm acc}
 \simeq
 25\,\eta Z\sigma^{1/2}
 \left(\frac{\beta L_{\rm k}}{10^{38}\,\rm erg/s}\right)^{1/2}\,{\rm PeV}.
 \label{eq:emax_lk}
\end{equation}
For electron acceleration, radiative cooling may be efficient and determine the maximum energy, which can be referred to Eq.~(\ref{eq:pwn_max_c}). Protons (and nuclei) suffer less radiative cooling unless the acceleration zone is very compact, so in most cases the available potential drop is the main constraint for hadron acceleration.}

\subsubsection{Large-scale emission from jet--environment interactions}
\label{sec:microquasar-large-scale}

Among the LHAASO sources, V4641~Sgr and GRS~1915+105 are particularly relevant to PeV particle acceleration because their spectra extend to, or approach, the PeV range and their gamma-ray morphology extends well beyond the central binary.

V4641~Sgr was first reported as a UHE gamma-ray source by HAWC, which found a highly elongated structure extending to above 200~TeV \cite{hawc2024v4641}. LHAASO independently detected an extended source with a characteristic size of about $0.33^\circ$ and a centroid displaced by about $0.19^\circ$ from the binary. Its spectrum extends to about 0.8~PeV without a clear cutoff \cite{lhaaso2025microquasars}. H.E.S.S. subsequently resolved the source at lower energies and found very hard and elongated emission in the $1-20$~TeV range, with an indication of two components on opposite sides of the binary \cite{hess2026v4641}. Combining H.E.S.S., HAWC, and LHAASO reveals an SED that peaks around 100~TeV. The large physical size, of order tens to $\sim100$~pc depending on the geometry, clearly separates the main gamma-ray-emitting region from the compact binary. 

The nature of the radiating particles remains uncertain. In a hadronic picture, protons are accelerated by the microquasar and then propagate through the surrounding medium before producing gamma rays. Anisotropic propagation along the Galactic magnetic field can naturally give an elongated morphology and has been proposed to explain the HAWC and LHAASO observations \cite{neronov2025v4641,kleimenov2026v4641}. However, dedicated H~I and CO observations used in the recent H.E.S.S. analysis do not reveal a sufficiently dense gas target, making a purely hadronic model energetically demanding \cite{hess2026v4641}. This is consistent with the location of V4641~Sgr which is about 500\,pc above the Galactic plane. A leptonic model remains attractive if the extended region has a weak magnetic field of only a few $\mu$G and particle transport is sufficiently fast. Shear acceleration at the boundary of the large-scale jet provides one possible way of continuously producing a hard electron population throughout the emitting region\cite{wan2026v4641}. In this case the observed extension traces the size of the acceleration zone itself rather than the distance over which electrons propagate from the black hole. An alternative leptonic picture has been proposed by Ref.\cite{bykov2026v4641}, in which the interaction of a narrow trans-relativistic jet with a quasi-spherical accretion-disk wind produces a system of strong shocks on a scale of $\sim 10\,$pc that can accelerate electrons to PeV energies. Recent deep X-ray observations provide further support for such a picture. Recent XMM-\textit{Newton} and Einstein Probe/FXT observations\cite{Chi2026,Tsuji2026} have revealed faint, extended X-ray structures spatially coincident with parts of the TeV emission. If these X-rays are synchrotron emission from the same electron population responsible for the TeV--PeV emission, their weak flux implies a magnetic field of only $\sim1$--$2\,\mu{\rm G}$, allowing PeV electrons to survive for $\sim$kyr and making a large-scale leptonic scenario plausible. Note that a hadronic origin of the highest-energy photons has not been excluded because the apparently large required energy budget can be alleviated if the long-term intrinsic accretion power of V4641~Sgr is much larger than suggested by its historical X-ray luminosity, as indicated by binary-evolution simulations \cite{liu2026v4641}. 

GRS~1915+105 shows a somewhat different picture for the origin of the UHE emission. The first LHAASO analysis found an extended source with a spectrum extending up to 600\,TeV with a size of about $0.28^\circ$, whose centroid is displaced by $\sim0.1^\circ$ from the binary \cite{lhaaso2025microquasars}. A more dedicated LHAASO analysis using one additional year of data \cite{lhaaso2026grs1915} finds that the spectrum gradually softens toward the PeV range but shows no sharp cutoff, while the source size decreases moderately with increasing energy. Radio and millimeter observations have identified structures tens of parsecs from the black hole that are possibly related to long-term jet--ISM interaction, consistent with the scale and direction of the gamma-ray emission\cite{rodriguez1998grs1915,chaty2001grs1915,tetarenko2018grs1915}. Deep MeerKAT observations of GRS~1915+105 imply a hydrogen density of approximately $100$--$160\,{\rm cm^{-3}}$ in the ambient medium of the jet-driven bow shock \cite{motta2025grs1915}. Such a dense environment provides an efficient target for escaped relativistic protons and allows a hadronic model to interpret the LHAASO observation\cite{khangulyan2024grs1915}. The difference between V4641~Sgr and GRS~1915+105 suggests that the appearance of a microquasar at UHE gamma-ray energies is determined not only by its ability to accelerate particles, but also by the environment where it is located.

\subsubsection{Binary-scale PeV emission from Cygnus~X-3}
\label{sec:cygx3}

Cygnus~X-3 is a compact binary of $\sim 0.02$\,au orbit scale and 4.8\,h period, containing a compact object and a Wolf--Rayet companion with an exceptionally dense wind and intense radiation field. \textit{Fermi}-LAT and AGILE detected transient GeV gamma rays from the system more than a decade ago, including orbital modulation and a close connection with radio and X-ray state changes \cite{abdo2009cygx3,tavani2009cygx3}. These observations had already shown that relativistic particles are accelerated close to the binary. Both anisotropic IC emission from electrons and hadronic interactions in the dense stellar environment have been considered \cite{dubus2010cygx3,Zdziarski2012,sahakyan2014cygx3}. On the other hand, long-term MAGIC observations found no significant signal between about 0.1 and 7~TeV, even when the data were selected according to the source state and orbital phase \cite{magic2026cygx3}.

LHAASO has revealed a point-like emission extending deep into the PeV range associated with Cygnus~X-3\cite{lhaaso2026cygx3}. After correction for Galactic absorption, the intrinsic hard spectrum extends from about 60~TeV to 3.7~PeV (Fig.~\ref{fig:microquasar}), and the source is detected above 1~PeV at a significance of $6.4\sigma$. More interestingly, the UHE flux varies on month timescales at high significance and becomes bright during GeV high states. LHAASO also reports a $3.2\sigma$ indication of modulation at the binary orbital period. Even though the orbital modulation still requires confirmation, the combination of variability and the orbital signal strongly implies that the PeV gamma rays are produced within the binary system.

Producing PeV gamma rays with electrons is difficult in such an environment. For a leptonic interpretation of the PeV emission,  electrons need to be accelerated to at least several PeV. The Hillas criterion then imposes a constraint on the magnetic field of the acceleration zone as $B>11(E/1\,{\rm PeV})(R/0.02\,{\rm au})^{-1}\beta_j^{-1}\,$G where $\beta_j$ is the jet velocity in units of the speed of light. Relativistic electrons suffer very rapid synchrotron losses in such a strong magnetic field and cannot be energized to PeV energies. Therefore, a hadronic origin is favored for the highest-energy component. Protons accelerated in the system can interact with the dense Wolf--Rayet wind through $pp$ collisions or radiation from the accretion disk through the photopion production, while at the highest energies photopion production with the stellar radiation field become important. The LHAASO spectrum can be reproduced if protons reach tens of PeV present at the binary scale \cite{lhaaso2026cygx3}. However, the specific acceleration mechanism and site of these very energetic protons are not clear yet.

GeV and PeV emissions are temporally correlated, so the GeV emission may provide clues to the acceleration and emission site of $\sim 10\,$PeV protons. Using contemporaneous \textit{Fermi}-LAT spectra and orbital modulation, \cite{zhang2026cygx3} constrained the GeV emitting region to a characteristic height of $2.8^{+2.1}_{-0.8}\times10^{11}$~cm along the jet. The inferred magnetic field--size product gives a maximum proton energy of only about 0.3~PeV there, well below the energy required by the LHAASO photons. The PeV accelerator must therefore lie farther inward, where a larger electromagnetic potential (or magnetic field--size product) is available \cite{zhang2026cygx3}. \cite{Zdziarski2026} proposed that nuclei accelerated near the jet base escape farther downstream and that most of the observed PeV emission is actually produced by the counterjet, whose longer path through the stellar radiation and wind fields naturally explains the phase of the observed orbital modulation, although their proposed jet geometry may not be consistent with that constrained by the GeV orbital modulation \cite{zhang2026cygx3}. Ref.\cite{Wan2026} suggested that the disk wind may dominate the trajectory of escaping protons and provided a self-consistent explanation to GeV and PeV spectral and orbital modulation simultaneously. Interestingly, this scenario may explain the extraordinary, yet unconfirmed, PeV flux from Cygnus~X-3 reported in 1980s by some experiments\cite{Samorski1983,LloydEvans1983} if the disk wind property was different at that time. In other multimessenger studies,
several possible particle acceleration sites and mechanisms have been explored \cite{kachelriess2025,wei2025microquasars}, showing that the PeV emission of Cygnus~X-3 can arise from either $p\gamma$ or $pp$ interactions depending on the location of the accelerator. Nevertheless, the detailed particle acceleration mechanism has not yet been established and needs further study.

Cygnus~X-3 demonstrates that protons can be accelerated beyond PeV energies on scales comparable to the binary orbit, very close to the central engine. Similar compact acceleration zones may also operate in systems such as V4641~Sgr, GRS~1915+105 and other microquasars detected by LHAASO, even though their observed UHE emission is produced much farther away. Cygnus~X-3 is special in the sense that it provides very favorable conditions for revealing such a small-scale hadronic accelerator, with its very compact orbit, luminous WR companion, dense stellar wind, all of which enhance the efficiency of $pp$ and $p\gamma$ interactions within the binary. In systems with wider orbits, less luminous companions, or weaker winds, PeV protons accelerated near the black hole may escape from the binary with little radiation and become visible only later, after interacting with the large-scale jet environment or surrounding interstellar medium. The difference between compact and extended gamma-ray sources may reflect not only where the highest-energy particles are accelerated, but also where suitable targets are available for converting their energy into observable gamma rays.

\subsection{The Cygnus Bubble}
\label{sec:cygnus-bubble}

The Cygnus Bubble differs from the sources discussed above in the nature of its environment and the possible accelerators embedded within it. Cygnus~X is one of the most active nearby star-forming complexes, containing the massive stellar association Cygnus~OB2, numerous young stars and compact objects, molecular clouds, and evidence for past supernova activity. Massive stellar winds can reach velocities well above 1000\,km/s and have been proposed as candidate accelerators of PeV protons \cite{casse1980,cesarsky1983,aharonian2019massivestars}.  {  The maximum achievable energy in this scenario follows the same constraint given by Eq.~(\ref{eq:emax_lk}).} The gamma-ray emission does not necessarily trace a single accelerator. Instead, it may result from the combined action of particle acceleration, escape, and interaction with the highly structured gas and radiation fields throughout the region.

\begin{figure}
    \centering
    \includegraphics[width=0.7\linewidth]{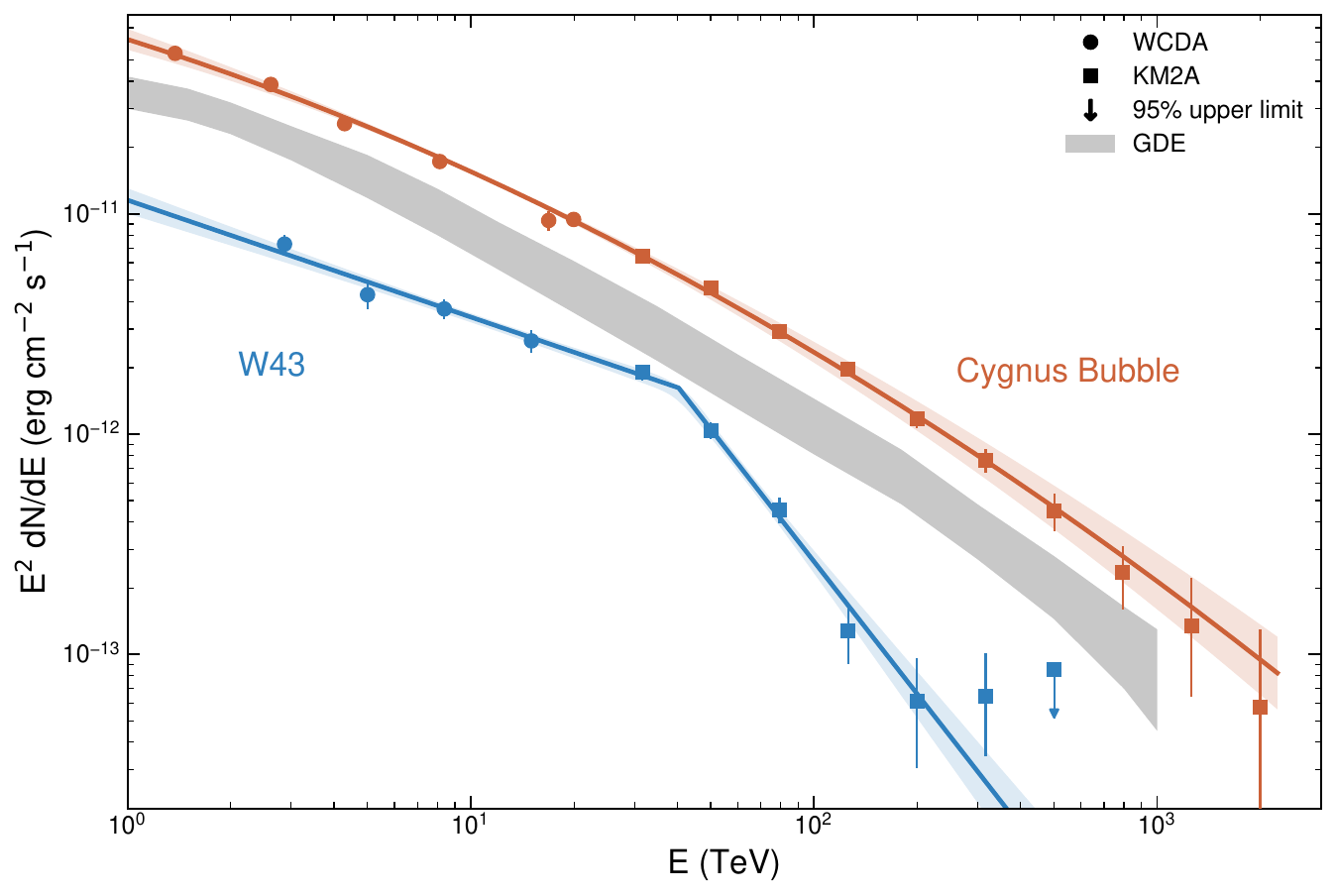}
    \caption{Spectral energy distributions of the Cygnus Bubble (orange) and W43 (blue) measured by LHAASO. Circles and squares denote WCDA and KM2A measurements, respectively, and the downward arrow indicates a 95\% confidence-level upper limit. For the Cygnus Bubble, the flux is integrated within $6^\circ$ of the bubble center including the contribution from the local Galactic diffuse emission (GDE). The orange curve and shaded region show the log-parabola fit and its uncertainty, while the grey band shows the ``GDE estimation~1'' shown in Ref.\protect\cite{lhaaso2024bubble}. The spectrum W43 is shown here for comparison, the spectral points are taken from Ref.\protect\cite{lhaaso2025w43}. The blue curve is a broken-power-law fit and the shaded region represents the corresponding $1\sigma$ uncertainty.}
    \label{fig:starcluster}
\end{figure}

The Cygnus region was first identified as an extended GeV ``cocoon'' by \textit{Fermi}-LAT, with subsequent observations by ARGO-YBJ and HAWC extending the spectrum into the TeV range \cite{ackermann2011cocoon,bartoli2014cygnus,abeysekara2021cygnus}. Even before the LHAASO measurement, the continuation of the Cocoon emission toward the PeV range had been discussed\cite{liu2021subpev,abeysekara2021cygnus}, in light of concentration of diffuse gamma-ray photons above 400\,TeV from the Cygnus direction reported by AS$\gamma$\cite{asgamma2021_diffse}. LHAASO revealed a much larger structure, the Cygnus Bubble, extending over more than $6^\circ$ and reaching photon energies of about 2~PeV \cite{lhaaso2024bubble} (Fig.~\ref{fig:starcluster}). The emission contains a broad diffuse component together with brighter structures around Cygnus~OB2 and hotspots correlated with dense molecular gas. If the highest-energy emission is hadronic, the observations require parent protons extending to at least several PeV and probably beyond 10~PeV. The large extent and the correlation with gas result from particle transport after acceleration.

An interesting comparison is provided by W43, another massive star-forming region detected by LHAASO as an extended VHE--UHE gamma-ray source \cite{lhaaso2025w43}. Its gamma-ray emission is also associated with a dense and energetic star-forming environment, but, unlike the gradual spectral extension of the Cygnus Bubble into the PeV range, W43 shows a much sharper spectral break (Fig.~\ref{fig:starcluster}). In a hadronic interpretation, the inferred proton spectrum breaks or cuts off at only a few $10^{14}$~eV \cite{lhaaso2025w43}. The contrast between these two star-forming regions raises an important question: what physical conditions determine the maximum particle energy that can be achieved in a massive star-forming environment?

Massive-star activity in Cygnus~OB2 is an obvious candidate for producing such extreme particles, but both the location and the physical mechanism of the acceleration remain debated. In one class of models, the winds of the massive stars merge into a collective cluster wind and particles are accelerated by diffusive shock acceleration at its termination shock. Under sufficiently strong turbulence, approaching the Kraichnan or Bohm regime, such a shock can accelerate protons into the PeV range and generally reproduce the observed high-energy emission \cite{morlino2021clusterwind,blasi2023cygob2,menchiari2024cygob2}. However, hydrodynamic simulations by \cite{vieu2024cygob2} suggest that because Cygnus~OB2 is a relatively loose association, its stellar winds may not merge into a coherent large-scale supersonic outflow and hence may never form the required collective termination shock. In this scenario, particle acceleration may instead occur at individual wind shocks or wind-wind interactions\cite{bykov2022}, in turbulent wind-interaction regions\cite{vink2024superbubbles}, or in supernova remnants evolving inside the superbubble\cite{haerer2025cygnus}. In particular, Ref.\cite{haerer2025cygnus} showed that a powerful supernova occurring about $5\times10^4$~yr ago could account for much of the 10~TeV--PeV emission, while electrons accelerated in stellar winds may contribute at lower energies.

The highest-energy part of the Bubble might have a different origin from the bulk of its emission. Following the discovery that Cygnus~X-3 accelerates hadrons to well beyond PeV energies, it has been proposed that most of these protons escape from the compact binary and subsequently diffuse through the surrounding ISM of $\sim 1\,$kpc\cite{lhaaso2026cygx3,Shi2026}. In this model, Cygnus~X-3 can make a substantial, and possibly dominant, contribution to the Cygnus Bubble above several hundred TeV, while the lower-energy emission remains primarily associated with the star-forming activity in Cygnus~OB2 with a maximum proton energy of a few hundred TeV. However, the power of escaping PeV protons from Cygnus~X-3 as revealed by detailed modeling of its UHE gamma-ray spectrum and orbital light curve\cite{Wan2026,Zdziarski2026} might not be sufficient to supply the bubble's flux, if the high-state duty cycle is not sufficiently high and the diffusion coefficient within the surrounding ISM is not sufficiently low.

Disentangling these possibilities will require spatially resolved multiwavelength and multimessenger observations. Improved GeV--PeV measurements can determine how the morphology changes with energy and test whether the highest-energy emission follows the same spatial component as the lower-energy cocoon or becomes increasingly concentrated toward compact accelerators such as Cygnus~X-3. X-ray observations are particularly important for constraining any competing high-energy electron population through its synchrotron emission, while radio, infrared, H~I, and molecular-line measurements can map the stellar winds, shocks, and target gas needed to connect the gamma-ray morphology with individual accelerators and CR propagation. Neutrinos provide the most direct test of the hadronic component. Existing IceCube searches have not yet found a significant counterpart to the Bubble \cite{li2024cygnusneutrino}, but more sensitive next-generation neutrino detectors could establish whether the TeV--PeV gamma rays are accompanied by the neutrino flux expected from hadronic interactions. Together with the comparison to regions such as W43, such observations may reveal not only which objects accelerate Galactic CRs to PeV energies, but also which environmental conditions allow a star-forming region to operate as a PeVatron.

\section{Implications for the Galactic PeV cosmic-ray population}
\label{sec:implications}

The observations reviewed above demonstrate that several classes of Galactic objects are capable of accelerating particles into the PeV regime. A more fundamental question, however, is whether the population of such extreme accelerators can supply the Galactic CRs observed around the knee. The answer cannot be straightforwardly inferred from the gamma-ray luminosities of the known PeV emitters. For leptonic sources, the proton power is largely unconstrained, while for hadronic sources the gamma-ray luminosity depends not only on the proton injection power but also on particle escape and the density of the target material. Nevertheless, model-dependent estimates can provide a useful indication of the PeV proton power of a given source. Defining the proton injection power per logarithmic energy interval as ${\cal L}_{p}(E_p)\equiv E_p^2\frac{d\dot N_p}{dE_p}$, the model that simultaneously reproduces the spectrum and surface-brightness profile of the Cygnus Bubble requires a total proton injection power of $\simeq1.1\times10^{37}\,{\rm erg\,s^{-1}}$ above 1~GeV, with
$d\dot N_p/dE_p\propto E_p^{-2.25}\exp(-E_p/5\,{\rm PeV})$ \cite{lhaaso2024bubble}. This corresponds to only
${\cal L}_{p}(1\,{\rm PeV})\simeq7\times10^{34}\,{\rm erg\,s^{-1}}$.
For the Crab Nebula, the spatially dependent lepto-hadronic model of \cite{liang2025pwn} found a present-day proton injection power at 1~PeV of ${\cal L}_p(1\,{\rm PeV})\simeq(2-3)\times10^{36}\, {\rm erg\,s^{-1}}$, depending only weakly on whether the injection index is 1.5 or 2.0. For Cygnus~X-3, the modeling of \cite{lhaaso2026cygx3} gives $\dot W_p(E_p>1\,{\rm PeV})\simeq5\times10^{36}(a/z_j)\,{\rm erg\,s^{-1}}$,
where $a$ is the orbital separation and $z_j$ characterizes the size of the emission region. 

On the other hand, the required Galactic CR power can be estimated in a simple leaky-box description as
\begin{equation}
    {\cal L}_{\rm CR}(E)
    \equiv E^2 Q_{\rm Gal}(E)
    \simeq
    \frac{V_{\rm h}\,\bar u_{\rm CR}(E)}
         {\tau_{\rm esc}(E)},
    \label{eq:crpower_general}
\end{equation}
where $V_{\rm h}$ is the effective CR confinement volume and
$\bar u_{\rm CR}$ is the Galactic-average CR energy density per logarithmic energy interval \cite{strong2007crprop}.
For a cylindrical diffusion halo with radius $R_{\rm G}$ and half-height $H$,
$V_{\rm h}\simeq2\pi R_{\rm G}^2H$.
If escape is dominated by diffusion, $\tau_{\rm esc}(E)\simeq\frac{H^2}{2D(E)}$, while the locally measured CR intensity gives $u_{\rm CR,\odot}(E)=(4\pi/c)E^2J(E)$.
Writing $\bar u_{\rm CR}=\xi_{\rm G}u_{\rm CR,\odot}$, where $\xi_{\rm G}$ describes how the average CR density in the Galaxy differs from that measured at Earth, one obtains
\begin{equation}
    {\cal L}_{\rm CR}(E)
    \simeq
    \frac{16\pi^2R_{\rm G}^2}{cH}
    D(E)\,\xi_{\rm G}E^2J(E).
    \label{eq:crpower_diffusion}
\end{equation}
Using the proton spectrum measured by LHAASO around 1~PeV,
$E^2J_p(E)\simeq1.5\times10^{-7}\, {\rm erg\,cm^{-2}\,s^{-1}\,sr^{-1}}$ \cite{lhaaso2025proton}, the required PeV proton injection power is approximately
\begin{equation}
\begin{split}
    {\cal L}_{\rm CR,p}(1\,{\rm PeV})
    \simeq &\ 10^{39}\ {\rm erg\,s^{-1}}\,
    \xi_{\rm G}
    \left(\frac{D_{\rm PeV}}
    {10^{31}\ {\rm cm^2\,s^{-1}}}\right)\\
    &\times
    \left(\frac{H}{5\,{\rm kpc}}\right)^{-1}
    \left(\frac{R_{\rm G}}{15\,{\rm kpc}}\right)^2
    \left[
    \frac{E^2J_p(E)}
    {1.5\times10^{-7}\,
    {\rm erg\,cm^{-2}\,s^{-1}\,sr^{-1}}}
    \right].
    \label{eq:pevpower}
\end{split}
\end{equation}
The numerical value depends on the diffusion coefficient $D$. Faster diffusion implies more rapid escape and hence requires a larger Galactic CR injection power, and vice versa. For example, extrapolating a conventional diffusion coefficient
\begin{equation}
    D(E)=4\times10^{28}
    \left(\frac{E}{1\,{\rm GeV}}\right)^{1/3}
    {\rm cm^2\,s^{-1}}
\end{equation}
to 1~PeV gives $D_{\rm PeV}\simeq4\times10^{30}\,{\rm cm^2\,s^{-1}}$ and hence
${\cal L}_{\rm CR,p}(1\,{\rm PeV})\simeq4\times10^{38}\xi_{\rm G}\,{\rm erg\,s^{-1}}$.
For $\xi_{\rm G}\simeq1$, this is equivalent to the PeV output of $\sim 100$ Cygnus~X-3-like accelerators operating in the Galaxy.

Currently, known potential hadronic PeVatrons with a proton power comparable to Cygnus~X-3 are rare. There may be a comparable number of such sources in the southern sky outside the LHAASO field of view, but it is not obvious that this number can account for the difference. One possibility is that the average Galactic diffusion coefficient at PeV rigidities is smaller than the simple Kolmogorov extrapolation used above. DAMPE measured a significant hardening of the B/C and B/O ratios at about $100$~GeV/n, with the B/C slope changing from approximately $-0.36$ below the break to $-0.20$ above it \cite{dampe2022bc}. In the simplest propagation interpretation, where B/C approximately traces $D^{-1}$, this corresponds to a decrease of the diffusion index from $\delta\simeq0.35$ to $\simeq0.2$. Extrapolating such a high-rigidity behavior to PeV energies would give $D_{\rm PeV}$ of order $10^{30}\,{\rm cm^2\,s^{-1}}$, about a factor of 4 below the unbroken $D\propto\mathcal R^{1/3}$ extrapolation, and would reduce the required PeVatron power the same factor. Moreover, the number of currently visible PeVatrons need not equal the number of sources contributing to the present CR sea. The residence time of CRs is
\begin{equation}
    \tau_{\rm esc}\simeq
    3.8\times10^6\ {\rm yr}
    \left(\frac{H}{5\,{\rm kpc}}\right)^2
    \left(\frac{D_{\rm PeV}}
    {10^{30}\ {\rm cm^2\,s^{-1}}}\right)^{-1}.
    \label{eq:pev_residence}
\end{equation}
Even for $D_{\rm PeV}=4\times10^{30}\,{\rm cm^2\,s^{-1}}$, extrapolated from the standard ISM diffusion coefficient, the residence time is close to $10^6$~yr. If the phase during which a particle accelerator operates as a very efficient PeVatron is much shorter than this residence time, a substantial fraction of the PeV CRs measured today may have been injected by accelerators that have already faded or entered a quiescent state. High-energy electrons accelerated at an early time have cooled and PeV protons accelerated at an early time have propagated far away from the source, so these accelerators may not appear as bright UHE gamma-ray sources today. In this case, the total Galactic CR power at PeV energies is time-dependent, and its value at the present time could be lower than the long-term average. 

In addition to the possible temporal variability in CR power, the spatial distribution could be inhomogeneous. In other words, the locally measured proton spectrum around the knee may not be representative of the Galactic average, corresponding to $\xi_{\rm G}<1$ in Eq.~(\ref{eq:crpower_diffusion}). If PeVatrons are sufficiently rare, the shorter residence time of PeV particles makes their spatial distribution more sensitive to the recent history of individual sources, and the Solar system may lie in a local enhancement produced by one or a few nearby accelerators. Ref.\cite{fang2026kneelocal} showed that such ``CR islands'' naturally become important when the Galactic PeVatron birth rate falls below about one percent of the supernova rate, and found that a source at a distance of order 1~kpc and an age of $\sim0.1-1\,$Myr could make a substantial contribution to the locally observed knee without producing an excessively bright present-day UHE gamma-ray halo. Ref.\cite{zhangyu2026remnants} have also suggested that anisotropic propagation of CRs can make the local PeV flux dominated by only a few nearby or recent microquasars. In such a scenario, using the locally measured PeV flux to infer a Galaxy-wide injection power would overestimate the true requirement.

CR anisotropy provides an independent test of these scenarios. In the isotropic diffusion approximation, the dipole anisotropy is
\begin{equation}
    \boldsymbol{\delta}(E)
    =-\frac{3D(E)}{c}
    \frac{\boldsymbol{\nabla}n_{\rm CR}(E)}
         {n_{\rm CR}(E)}.
    \label{eq:dipole}
\end{equation}
Measurements show that the CR distribution remains largely isotropic in the PeV regime, with large-scale amplitudes of only a few $10^{-3}$ \cite{aartsen2013anisotropy,abbasi2025anisotropy}. A nearby source that dominates the measured PeV flux could lead to too large a density gradient. The expected gradient depends strongly on the injection history. Two useful limiting cases are instantaneous burst-like injection and continuous injection at a constant rate:
\begin{equation}
n_s(E,r,t)=
\begin{cases}
\displaystyle
\frac{Q_b(E)}
     {[4\pi D(E)t]^{3/2}}
\exp\left[-\frac{r^2}{4D(E)t}\right],
& {\rm burst},\\[12pt]
\displaystyle
\frac{\dot Q_c(E)}
     {4\pi D(E)r}
\,{\rm erfc}\left[
\frac{r}{2\sqrt{D(E)t}}
\right],
& {\rm continuous},
\end{cases}
\label{eq:source_density}
\end{equation}
where $Q_b$ is the spectrum of all particles injected in the burst-like injection case and $\dot{Q}_c$ is the differential injection rate of particles in the continuous injection case. The corresponding logarithmic density gradients are
\begin{equation}
\frac{|\nabla n_s|}{n_s}
=
\begin{cases}
\displaystyle
\frac{r}{2D(E)t},
& {\rm burst},\\[10pt]
\displaystyle
\frac{1}{r}
+\frac{\exp(-x^2)}
{\sqrt{\pi D(E)t}\,{\rm erfc}(x)},
& {\rm continuous},
\end{cases}
\label{eq:source_gradient}
\end{equation}
with $x\equiv\frac{r}{2\sqrt{D(E)t}}$. For an old continuous source satisfying
$t\gg r^2/4D$, the second expression approaches
$|\nabla n_c|/n_c\simeq1/r$. If the source contributes a fraction
$f_s(E)=n_s(E)/n_{\rm CR}(E)$
of the locally observed CR density, its contribution to the dipole is therefore approximately
\begin{equation}
\delta_s(E)\simeq
\begin{cases}
\displaystyle
f_s\frac{3r}{2ct},
& {\rm burst},\\[10pt]
\displaystyle
f_s\frac{3D(E)}{cr},
& {\rm continuous,\ quasi\mbox{-}steady}.
\end{cases}
\label{eq:source_dipole}
\end{equation}
For burst-like injection, the CR cloud becomes progressively smoother as it expands, and the explicit dependence on $D$ cancels from the dipole for a fixed local source fraction. For continuous injection, the source maintains an approximately $n\propto r^{-1}$ profile and the anisotropy increases directly with $D$. Numerically,
\begin{equation}
    \delta_b\simeq
    4.9\times10^{-2}f_s
    \left(\frac{r}{1\,{\rm kpc}}\right)
    \left(\frac{t}{10^5\,{\rm yr}}\right)^{-1},
\end{equation}
while in the quasi-steady limit
\begin{equation}
    \delta_c\simeq
    0.13 f_s
    \left(\frac{D_{\rm PeV}}
    {4\times10^{30}\,{\rm cm^2\,s^{-1}}}\right)
    \left(\frac{r}{1\,{\rm kpc}}\right)^{-1}.
\end{equation}
Thus, for example, a $10^5$-yr-old burst source at 1~kpc contributing ten percent of the local PeV flux would already produce a dipole of order $5\times10^{-3}$ in this simple picture. A continuously injecting source can be even more strongly constrained for a large diffusion coefficient. Sources with a finite activity period followed by a fading phase should lie between these two limiting cases.

We can see that the diffusion coefficient affects the Galactic power budget and the anisotropy constraint in the same direction. A larger $D_{\rm PeV}$ increases the total injection power required to maintain the Galactic PeV population [Eq.~(\ref{eq:crpower_diffusion})], while increasing the anisotropy produced by a nearby continuously injecting source [Eq.~(\ref{eq:source_dipole})] at the same time. If Galactic PeVatrons are fewer in number, the contribution of each individual source becomes larger, which makes the imprint of the nearest source on the local spectrum and anisotropy more difficult to hide. This provides an important consistency test of models in which the knee is dominated by a small number of nearby accelerators.
The above estimate assumes isotropic diffusion. In reality, the observed CR dipole anisotropy can also be strongly affected by the local magnetic-field configuration. For anisotropic diffusion, the dipole vector is more generally written as
\begin{equation}
\boldsymbol{\delta}
=
\frac{3}{c}
\frac{\mathbf{D}\cdot\nabla n}{n},
\end{equation}
where $\mathbf{D}$ is the diffusion tensor. If diffusion is much faster along the ordered magnetic field than across it, the observed dipole tends to be projected along the local field direction rather than simply tracing the CR density gradient. Consequently, the dipole amplitude produced by a nearby source may be either suppressed or enhanced depending on the relative orientation between the source-induced density gradient and the local magnetic field. Local magnetic turbulence and structures such as the Local Bubble may further modify both the amplitude and phase of the observed anisotropy. 

On the other hand, we should also note that Galactic CR transport is not necessarily described by standard diffusion on all relevant scales. Equation~(\ref{eq:dipole}) assumes the Fickian relation $\boldsymbol{j}=-\mathbf{D}\cdot\nabla n_{\rm CR}$, whereas magnetic focusing and mirroring, field-line wandering, and the transition between ballistic and diffusive propagation can modify both the local CR gradient and the distribution of arrival directions. These effects may be particularly important for nearby PeV sources, for which the scattering mean free path can become comparable to the relevant spatial scales. In addition, different source contributions can also partially cancel as vectors \cite{mertsch2015anisotropy,ahlers2016anisotropy}. LHAASO has shown that the proton and helium spectra evolve differently in this energy range. Protons overtake helium at about 0.7\,PeV, while helium becomes dominant again at about 5\,PeV, with the two components showing distinct hardenings and knees \cite{lhaaso2025proton,lhaaso2026helium}. Such a strongly energy-dependent p/He ratio is difficult to describe with a single source population having a fixed elemental composition, and may instead indicate the superposition of different Galactic CR source populations with different compositions, spectra, or maximum rigidities \cite{yuan2026multisource}. If the proton and helium components around the knee are contributed by different source populations, they may have not only different diffusion coefficients but, more importantly, different spatial density gradients and dipole directions. The observed all-particle anisotropy would then be the vector sum of several components, making the interpretation of the spectrum and anisotropy more complicated.

\section{Summary}\label{conclusions}

Detection of PeV gamma-ray emitters by LHAASO has changed the study of Galactic PeVatrons from a search for candidate sources into detailed studies and modeling of specific sources. The Crab Nebula and the Aquila Booster (PSR~J1849-0001) show that pulsar winds can accelerate electrons to PeV energies with extreme efficiency. In particular, the relatively low spindown power of PSR~J1849-0001 and the low magnetic field of its nebula imply that the acceleration zone of PeV electrons may be upstream of the termination shock. Microquasars provide perhaps the clearest demonstration that PeV particle acceleration can occur on multiple spatial scales. V4641~Sgr and GRS~1915+105 show extended UHE emission associated with the large-scale jets, while Cygnus~X-3 demonstrates that particles can reach PeV energies on scales comparable to the binary orbit. These cases should not be regarded as mutually exclusive acceleration modes. The Cygnus Bubble brings us to a different environment of complex star-forming activities containing massive stars, past supernova activity, compact objects, dense gas, and several possible accelerators. The origin of its highest-energy particles remains unsettled. Collective stellar-wind acceleration can reach the PeV regime under favorable conditions, although whether Cygnus~OB2 actually forms the required large-scale wind termination shock remains debated. 

Although this review focuses on PeV gamma-ray emitters, sources whose spectra soften or cut off well below the PeV range are also important for understanding the origin of PeV CRs. A source may be efficient PeV CR accelerator even if its spectrum does not extend to PeV, such as SS~433. Furthermore, even for those sources with identified spectral cutoff well below PeV, they provide a control sample against which the exceptional PeV gamma-ray emitters can be compared within the same source class. Examples include the PWN of PSR~J1740+1000 (LHAASO~J1740+1000), the star-forming region W43 (LHAASO~J1848-0153), and Cygnus~X-1 (LHAASO~J1958+3522) and so on, whose spectra reach the UHE regime but soften or terminate substantially earlier than the most extreme PeV emitters \cite{lhaaso2025_j1740,lhaaso2025w43,lhaaso2025microquasars}. Such comparisons may be valuable for identifying which physical conditions control the maximum particle energy (e.g., the available electromagnetic power, energy-dissipation mechanism, magnetic-field strength and geometry, turbulence level, source age, and activity history). In this sense, sources that fail to reach the PeV regime are not simply less interesting accelerators. Instead, together with the PeV emitters, they define the parameter space needed to understand what makes an astrophysical accelerator efficient enough to reach the Galactic CR knee. Also, we need to note that an earlier gamma-ray spectral cutoff does not always correspond to a less efficient accelerator, because the gamma-ray cutoff can also be influenced by radiative losses, target fields, or gamma-ray absorption.

In the future, enlarging the sample of Galactic PeV gamma-ray emitters will remain one of the most important steps toward understanding the origin of PeV CRs. The continuously increasing exposure of LHAASO will reveal fainter sources and improve the statistics of known emitters, while the planned Southern Wide-field Gamma-ray Observatory (SWGO) will provide complementary coverage of the southern sky \cite{swgo2025prospects}. A larger and more complete sample will make it possible to compare extreme accelerators with less efficient members of the same source classes and help identify the physical conditions required to reach the PeV regime. Higher photon statistics for bright sources will also provide more precise spectra, allowing searches for additional emission components and, in particular, measurements of the detailed spectral shape in the cutoff regime. The cutoff shape carries information about both the radiation mechanism and the underlying particle distribution, which in turn provides constraints on the particle-acceleration process\cite{zirakashvili2007,lefa2012}. Next-generation IACTs such as CTAO, LACT, and ASTRI will resolve gamma-ray morphologies and their energy dependence with much greater precision, helping to distinguish acceleration sites from regions illuminated by particles after propagation. Multiwavelength and multimessenger observations will be equally important for determining the nature and location of the accelerated particles. X-ray measurements can constrain the highest-energy electron population and the magnetic field, while CO and H~I line observations can map the target gas needed to assess hadronic scenarios and CR propagation in the surrounding environment. Next-generation neutrino telescopes will provide a more direct test of PeV hadron acceleration. Eventually, the origin of Galactic CRs will be clarified when we understand not only which objects accelerate particles to high energies, but also how the acceleration operates, how the particles escape and propagate, and why these processes collectively produce the CR spectrum, composition, and anisotropy as observed.

\bibliographystyle{ws-mpla}
\bibliography{ws-mpla}

\end{document}